\documentclass[aps,prd,superscriptaddress,showpacs,showkeys]{revtex4}
\begin{document}
\preprint{}
\title{Fundamental Matter and the
Deconfining Phase Transition in 2+1 D}
\author{Gerald V. Dunne}
\affiliation{Department of Physics, University of Connecticut,
Storrs, CT 06269-3046, USA}
\author{Alex Kovner}
\affiliation{Department of Physics, University of Connecticut,
Storrs, CT 06269-3046, USA}
\affiliation{Department of Mathematics and Statistics, University
of Plymouth,  Drake Circus, Plymouth,
PL4 8AA, UK}
\author{Shinsuke M. Nishigaki}
\affiliation{Department of Physics, University of Connecticut,
Storrs, CT 06269-3046, USA}
\date{July 5, 2002;  \ Revised: August 30, 2002}
\begin{abstract}
We analyze the effect of heavy fundamentally charged particles on the
finite temperature deconfining phase transition in the
2+1 dimensional Georgi-Glashow model. We show that in the presence of
fundamental matter the transition turns into a crossover. The
near critical theory is mapped
onto the 2 dimensional Ising model in an external magnetic field.
Using this mapping we determine the width of the crossover region as well
as the specific heat as a function of the
fundamental mass.
\end{abstract}
\maketitle

Recently, significant progress has been made in understanding
the deconfining  phase transition in 2+1 dimensions. The transition in
the
$SU(2)$ Georgi-Glashow model has been analysed in detail:
the order of the phase transition as well as the universality class have
been established explicitly without recourse to universality
arguments,
and the dynamics of the phase transition was given a simple
interpretation
in terms of the restoration of magnetic symmetry \cite{gg1}.
In subsequent work, the effects of instantons at high temperature have
been  understood, the dynamics of the deconfining transition
has been related to the properties of the confining strings, and
the analysis has also been extended to $SU(N)$ gauge theories
at arbitrary $N$ \cite{gg2}. The effects
of the variability of the Higgs field mass were studied in
\cite{antonov}, and some analogies between the
mechanism of the deconfining transition in 2+1 dimensions and chiral
symmetry restoration in QCD have been suggested \cite{gg3}.
These results have recently been reviewed and summarized in \cite{kk}.
Also, an interesting interpretation of these results has recently been
given \cite{mike} in the context of the Svetitsky-Yaffe conjecture
\cite{sy} and the general role of center symmetry in abelian projection.

In this Letter we ask how the properties of the transition change in the
presence of  dynamical particles in the fundamental representation of the
gauge group. We consider the $SU(2)$ Georgi-Glashow model in the
presence of a heavy fundamental field, which we take to be a scalar
(being heavy, similar results should hold for fundamental fermions 
\cite{note}). The
Lagrangian of the theory is
\begin{eqnarray}
{\mathcal L}= -{1\over 2g^2} \mbox{tr}\left(F_{\mu \nu}F^{\mu
\nu}\right)
+ {1\over 2}(D^{ab}_\mu h^b)^2 -{\lambda
\over  4}(h^a h^a - v^2)^2
+|D^{\alpha\beta}_\mu \Phi^\beta|^2 -M^2\Phi^*\Phi .
\label{model1}
\end{eqnarray}
Here $h^a$ is the Higgs field in the adjoint representation, and $\Phi$
is
a  scalar field in the fundamental representation of $SU(2)$.

We will be interested throughout this paper in the weakly coupled
regime $g^2 \ll v$.
In this regime, perturbatively the gauge group is
broken to $U(1)$ by the large expectation value, $v$, of the Higgs field.
The Higgs and the two gauge bosons $W^\pm$ are heavy with masses $
M^2_H= 2\lambda v^2$, and $M^2_W=g^2v^2$.
We take the fundamental field $\Phi$ to be much heavier than the
charged $W$-bosons :
\begin{equation}
M^2\gg M^2_W.
\end{equation}

Perturbatively the theory behaves very much like 2+1 dimensional
electrodynamics with spin one charged matter.
However, nonperturbative effects are very important at large
distances. As shown by Polyakov \cite{Polyakov}, their effect is that the
photon, which is perturbatively massless, acquires a finite (but
exponentially small) mass and the charged $W^\pm$ become linearly
confined
at large distances with nonperturbatively small string tension.

Let us first summarize what is known \cite{gg1,kk} about this theory
{\it without} the heavy fundamental matter field $\Phi$.
At zero temperature confinement is a consequence of the spontaneous
breaking of the magnetic $Z_2$ symmetry \cite{thooft,kovner}. The $Z_2$
symmetry transformation is generated by the Wilson loop along the spatial
boundary of the system
\begin{eqnarray}W(C \rightarrow \infty ) = \mbox{exp}\left( {i\over
2}\int d^2 x
B(x)\right),
\end{eqnarray}The order parameter $V(x)$ for this $Z_2$ transformation
is the operator
that creates an elementary magnetic vortex of flux $2\pi/g$
\begin{eqnarray}V(x) = \mbox{exp}\left({{2\pi i\over g}} \int_C
dx^i \epsilon_{ij}{h^a\over
|h|} E^a_j(x)\right).
\label{v}
\end{eqnarray}
Here, $C$ is a contour beginning at $x$ and going to
spatial infinity.
Despite the appearance of the contour in the definition, this operator
$V(x)$ is in fact local, gauge invariant and Lorentz scalar
\cite{kovner}.
The action of the spatial Wilson loop on $V(x)$ is given by
\begin{eqnarray}
W(C \rightarrow \infty )V(x)W^\dagger(C \rightarrow\infty) =- V(x)
\end{eqnarray}
which is a realization of the 't Hooft algebra \cite{thooft,kovner}.
At low energies the theory  is described by the $Z_2$
invariant Lagrangian of the vortex field $V$
\begin{equation}
{\mathcal L}_{\rm eff}= \partial_\mu V\, \partial^\mu V^* -
\lambda \bigl( V\, V^* - {g^2\over 8\pi^2}\bigr)^2 +
{m_{ph}^2\over 4} \left\{ V^2 + (V^*)^2\right\}.
\label{lowlagrangian}
\end{equation}
with the photon mass \cite{Prasad}
\begin{eqnarray}m_{ph}^2= {16 \pi^2 \xi\over{g^2}},
\ \ \
  \xi = \mbox{constant}
{M_W^{7/2}\over g} e^{-{4\pi M_W \over g^2} }
\label{mph}
\end{eqnarray}
and the vortex self-coupling $\lambda={2\pi^2M_W^2\over g^2}$. At weak
coupling, $g^2\ll v$,  the radial degree of freedom of $V$ is very heavy,
and is practically frozen. Thus, the effective Lagrangian reduces to an
effective Lagrangian for the phase of $V$:
\begin{eqnarray}\L_{eff} =
  {g^2\over 8 \pi^2}(\partial_{\mu}\chi)^2 + {m_{ph}^2 g^2\over
16\pi^2}\cos
2\chi,
\label{sine-Gordon}
\end{eqnarray}
where
\begin{eqnarray}V={g\over {\sqrt 8}\pi}\exp{i\chi} .
\end{eqnarray}The effective Lagrangian (\ref{sine-Gordon}) for
slowly varying fields $\chi$ is equivalent to the one derived
by Polyakov \cite{Polyakov} from the monopole (more precisely,
monopole-instanton) plasma picture, with
$\xi= {m_{ph}^2 g^2\over 16\pi^2}$ being the monopole fugacity.
The only additional information contained in (\ref{lowlagrangian}) is
that
the field $\chi$ should be treated as a phase. Thus rough configurations,
where $\chi$ changes by $2\pi$ between adjacent points in space have
finite energy. This is  important, since the charged particles, $W^\pm$
in this  representation show up as solitons of the
field $V$ with unit winding number, or vortices of the phase $\chi$
\cite{kovner}.
In these configurations $\chi$ is indeed discontinuous along
some cut, but the cut itself does not cost energy.

As shown in \cite{gg1}, even though the $W^\pm$ bosons are heavy, they
cannot be neglected at finite temperature. Their presence determines the
properties of the deconfining phase transition. The physics of the phase
transition is the following. At finite temperature the thermal ensemble
is
populated by  $W$ bosons, with density  proportional to their fugacity
\begin{eqnarray}\mu\propto e^{-M_W\over T} .
\label{mu}
\end{eqnarray}The $W$ bosons interact via a confining potential
\cite{Polyakov}.
Each
$W^+$ boson is a source of {\it two} confining strings, which both end on
the same nearby $W^-$ boson, and so at low temperature
the $W$ bosons are bound in pairs.

However, this confining interaction is weak. The
width of the confining string is proportional to the inverse mass $m_{\rm
ph}$ of the photon, given in (\ref{mph}), and is therefore very large.
When the average distance between the $W$ bosons becomes the same as the
width of the string, the confining interaction becomes irrelevant.
The two confining strings emanating from a given
$W^+$ do not have to end on the
same $W^-$ boson any longer, but rather the strings form a percolating
network.
The individual $W$ bosons therefore  have no memory of
their nearest neighbours, and
are free to wander independently in the thermal vacuum, thereby
forming a charged plasma.
Since the $W$ bosons are vortices of the phase $\chi$, in the plasma
state
the phase $\chi$ is disordered, and thus the magnetic $Z_2$ symmetry is
restored \cite{ka}. The transition happens at the point where the
fugacity
(\ref{mu}) of the $W$ bosons becomes equal to the fugacity (\ref{mph}) of
monopoles,
\begin{eqnarray}
T_c={g^2\over 4\pi} .
\end{eqnarray}
To analyze the phase transition quantitatively,  note that the
critical temperature is much larger than
the mass of the photon, and thus dimensional reduction is valid in the
critical region. The dimensionally reduced theory in addition
to the terms present in (\ref{sine-Gordon}) contains
contributions due to the finite density of $W$ bosons \cite{gg1}.
Thus, the two dimensional Euclidean Lagrangian that describes the
transition  region is
\begin{eqnarray}
\L = { g^2 \over 8 \pi^2 T } (\partial_{\mu}\chi)^2 +
\zeta\cos 2 \chi
+ \mu \cos \tilde{\chi}
\label{sinevortex}
\end{eqnarray}
where $\zeta$ is related to the monopole fugacity by $\zeta=\xi/T$, and
$\tilde \chi$ is the field dual to $\chi$,
\begin{eqnarray}
i\partial_{\mu}\tilde\chi= {g^2\over 2\pi T}
\epsilon_{\mu\nu}\partial^\nu \chi.
\label{chid}
\end{eqnarray}
As explained in \cite{gg1}, the dual field $\tilde \chi$ is directly
related to the zeroth component of the Abelian vector potential,
corresponding to the unbroken $U(1)$ gauge group in the original
formulation of the
Georgi-Glashow model, $\tilde\chi=2g\beta A_0$.
Thus the last term in eq.(\ref{sinevortex}) is nothing but the
potential $P^2+h.c.$ for the fundamental Polyakov line
\begin{eqnarray}
P=\exp\left({i\over 2}\,\tilde\chi\right)
\label{polyakovline}
\end{eqnarray}
which is indeed the leading contribution to the free energy due to heavy
charged particles.

The critical temperature $T_c=g^2/4\pi$
is special for three reasons. First, this is the point at which the
operators $\cos 2\chi$ and $\cos \tilde{\chi}$ have the same scaling
dimension, equal to one. Second, at this point the fields
$\chi\pm\frac{\tilde{\chi}}{2}$ become chiral (antichiral), as can be
seen from (\ref{chid}) :
\begin{eqnarray}
(\partial_1\pm i\partial_2)\left(\chi\pm \frac{\tilde{\chi}}{2}\right)=0.
\label{chiral}
\end{eqnarray}
Third, at
$T_c$ the coefficients of the two ``interaction terms'' in
(\ref{sinevortex}) become equal, $\zeta=\mu$. These facts all imply that
the theory can be conveniently fermionized by using
the standard bosonization/fermionization techniques \cite{Nersesyan}.
Defining the chiral and antichiral fermionic fields
\begin{equation}
\psi_{R}=a^{-1/2}\frac{1}{\sqrt{2}}\exp\left[ i\left(\chi+
{\tilde\chi\over 2}\right)\right]\quad ,\quad
\psi_{L}=a^{-1/2}\frac{1}{\sqrt{2}}\exp\left[- i\left(\chi-
{\tilde\chi\over 2}\right)\right]
\end{equation}
the potential terms in (\ref{sinevortex}) become
\begin{eqnarray}
a^{-1}\cos 2\chi=\psi_R^\dagger\psi_L -\psi_L^\dagger\psi_R\quad
,\quad 
a^{-1}\cos\tilde\chi=\psi_R^\dagger\psi^\dagger_L-\psi_L\psi_R.
\end{eqnarray}
The dimensional constant $a$ plays the role of the UV cutoff in
the effective theory, and is of the order of $T$ \cite{zarembo}.
Defining the Majorana fermions In terms of the Majorana fermions
\begin{eqnarray}
\rho={\psi+\psi^\dagger\over \sqrt 2}, \ \ \
\sigma={\psi-\psi^\dagger\over i \sqrt 2}
\end{eqnarray}
the effective Lagrangian (\ref{sinevortex}) becomes
\begin{equation}
L=\frac12\bar\rho\gamma_\mu\partial_\mu\rho+
\frac12 \bar\sigma\gamma_\mu\partial_\mu\sigma+
i\frac{\zeta+\mu}{2}\rho^T \gamma_2 \rho+
i\frac{\zeta-\mu}{2}\sigma^T\gamma_2\sigma
\label{fcrit}
\end{equation}
where the gamma matrices are taken as the
Pauli matrices : $\gamma_\mu=\tau_\mu$, and
$\bar{\psi}=\psi^\dagger \gamma_1$. Note that the
two Majorana fermions $\rho$ and $\sigma$ in (\ref{fcrit}) do not
interact with one another. The fermion $\rho$ has
Majorana mass of the order of the photon mass, while the fermion
$\sigma$ is
massless at criticality. At large distances ($d\gg \zeta^{-1}$) the
massive fermion decouples,  and so the long distance physics at
criticality  is governed by the theory of one
massless Majorana fermion, which describes the critical point of a single
2D Ising model. The implications of this for the deconfining phase
transition are analyzed in
\cite{gg1}.

We now ask how this picture changes in the presence of the heavy
fundamentally charged  matter field $\Phi$.
First, we note that when $M^2$ is finite, the nature of the magnetic
symmetry
changes. As discussed in \cite{fosco} it becomes a local rather than a
global
symmetry. The operator $V$ in (\ref{v}) is no longer a local operator.
The operator $V^2$ is still local, but it is not an order parameter for
$Z_2$. Thus there is no local order parameter that can distinguish
between
broken and unbroken magnetic symmetry. In this situation we do not expect
the  deconfining transition to remain second order. It should either
become first order with finite latent heat, or disappear altogether
into a
sharp but analytic crossover.

The way the fundamental matter affects the physics of the transition can
be understood
qualitatively from the following simple consideration.
For large $M^2$ the fugacity of the fundamental $\Phi$ particles is very
small, and thus they are present in the thermal ensemble
with very low density, proportional to
\begin{eqnarray}h=e^{-\frac{M}{T}} .
\end{eqnarray}As discussed earlier, below $T=T_c$, the ensemble of $W$
bosons
consists of dipoles bound together by a {\it pair} of confining strings.
A single $\Phi$ particle is a source (or a sink) of only one confining
string.
Any extra $\Phi$ particles in this ensemble therefore
have to bind in pairs between themselves. Thus, below the transition the
fundamental particles form an extra component of the ``dipole plasma'',
which does not
mix with the dominant (higher density) component consisting of $W$'s.
On the other hand, above the transition the strings emanating from $W$
bosons percolate through the whole ensemble, rather than ending on a
nearest particle.  In this situation, a $\Phi$ particle loses all memory
of any other $\Phi$ particles in its neighbourhood, since its own
confining string can easily end on a neighbouring $W$ boson, whose
density
is much higher. This change of the distribution of $\Phi$ particles
clearly leads to an increase of the entropy in the system.
Below the transition the contribution of the $\Phi$ particles to the
entropy can be estimated from
considering the thermal ensemble as an ensemble of ``dipoles''
with fugacity $h^2$
\begin{eqnarray}
\exp\,(TS_\Phi)=\sum_{n=0}^{\infty}(xh^2Va^{-2})^n{1\over
n!}=\exp\left( xh^2a^{-2}V\right)
\end{eqnarray}
where the factor of volume arises from an independent integration over
the center of mass coordinates of the dipoles. Also $x$ is a number of
order unity, encoding the fact that the fugacity of the dipole is not
exactly $h^2$ due to the interaction between the two particles forming
the dipole, and that there is an extra ``renormalization'' due to  the
integration over the internal states of the dipole. Thus parametrically
below the transition
\begin{eqnarray}
S_<\propto h^2 .
\label{entropybelow}
\end{eqnarray}
On the other hand, above the transition one can consider the
$\Phi$ particles as noninteracting and randomly distributed.  The result
for the entropy is then
\begin{eqnarray}
S_>\propto h .
\label{entropyabove}
\end{eqnarray}
Since $h$ is small, this means that the entropy rises strongly in
the transition region.   These estimates (\ref{entropybelow}) and
(\ref{entropyabove}) of the entropy are only valid far enough from the
transition, where $S$ can be expanded in powers of $h$. Thus, this simple
consideration is not sufficient to determine whether the change from
$S_<$
to $S_>$ takes place abruptly at some value of temperature
(a first order transition),  or smoothly over a finite range of
temperatures $\Delta T$ (a smooth crossover).
To probe this question more precisely we must determine how the presence
of
the heavy fundamental $\Phi$ particles modifies the dimensionally reduced
Lagrangian (\ref{sinevortex}) close to criticality.

First, it is clear that the presence of the heavy $\Phi$ particles does
not
change the self-interaction of the light photon, just as
the presence of heavy $W$'s does not.
However, in the presence of the fundamentally charged field $\Phi$, the
low
energy theory (\ref{sine-Gordon}) must now admit solitons with half the
topological charge. Thus the field
$\chi$  now has period $\pi$ rather than $2\pi$.
In addition, at finite temperature the presence of the $\Phi$
particles induces  a new term, similar to the last term
in (\ref{sinevortex}), but  with twice the periodicity and with a
coefficient proportional to the fugacity, $h$, of the $\Phi$ particles.
This contribution is proportional to the first power of the
Polyakov line in (\ref{polyakovline}). The derivation is completely
analogous to that  presented in \cite{gg1} and we will not discuss it in
detail. The dimensionally reduced Lagrangian therefore is
\begin{eqnarray}
\L = { g^2 \over 8 \pi^2 T } (\partial_{\mu}\chi)^2 +
\zeta\cos 2 \chi
+ \mu \cos \tilde{\chi}   +h\cos {\tilde{\chi}\over 2}    .
\label{sinevortex1}
\end{eqnarray}
Close to the transition temperature we can again fermionize the theory.
Expanding to leading order in $T-T_c$, and keeping only relevant
terms we find
\begin{equation}
L=\frac12\bar\rho\gamma_\mu\partial_\mu\rho+
i\zeta{\rho}^T\gamma_2\rho
+\frac12\bar\sigma\gamma_\mu\partial_\mu\sigma
-i\frac{\tau}{2}{\sigma}^T\gamma_2\sigma +h\Sigma
\label{lcrit}
\end{equation}
with
\begin{eqnarray}
\tau=
\left(T-{g^2\over 4\pi}\right){16\pi^2M_W\over g^4 }\zeta .
\label{taueq}
\end{eqnarray}
Since the fugacity $h$ of the heavy fundamental fields is small, we can
treat the term $h\Sigma$
as a perturbation. Recall
from  (\ref{fcrit}) that the system {\it without} this perturbation is
that of two decoupled 2D Ising models,
one of them close to criticality and another far away from criticality.
The term $h\cos {\tilde{\chi}\over 2}$ makes the two Ising models
coupled, resulting in the so-called Baxter--Ashkin-Teller model
\cite{ogilvie}. In his translation table between the sine-Gordon operators
and the Baxter operators,
Ogilvie \cite{ogilvie} has identified the operator $\cos
{\tilde{\chi}\over 2}$
(mass dimension $\frac14$)
with a product of a spin operator of one Ising model (A) and a disorder
operator
of another Ising model (B) : $\cos
{\tilde{\chi}\over 2}\leftrightarrow\hat{\sigma}^{(A)}\hat{\mu}^{(B)}$.
The conformal dimensions of both spin and disorder operators are each
$\frac{1}{16}$.
The operators $\cos {2{\chi}}\pm\cos {\tilde{\chi}}$ (mass
dimension $1$)
are identified with the energy (mass) operators
$\hat{\varepsilon}^{(A,B)}$, each having conformal
dimension=$\frac{1}{2}$, of  the Ising model A and B, respectively.
In the regime we are interested in, the Ising model A is deeply into the
ordered phase, so
the operator $\hat{\sigma}^{(A)}\hat{\mu}^{(B)}$ can be substituted by
$\hat{\mu}^{(B)}$ with mass dimension $\frac18$.
An alternative interpretation of this result is to integrate out the heavy
Majorana fermion. Its effect in the re-bosonized theory is
the placement of a background charge at infinity that enforces
$c=\frac{1}{2}$,
and a multiplicative renormalization of the sine-Gordon field.
Then the vertex operators $e^{i \tilde{\chi}}$ and $e^{i\tilde{\chi}/2}$
with the rescaled $\tilde{\chi}$ have
conformal dimensions $\frac{1}{2}$ and $\frac{1}{16}$ in the presence
of the background charge, respectively.

Either way, we interpret the effective Lagrangian (\ref{sinevortex1})
near criticality, and with heavy  fundamental matter fields, as that of
the $c=\frac{1}{2}$ conformal field theory of a Majorana fermion
with two perturbations, one of conformal dimension
$\frac{1}{2}$ and another one of conformal dimension $\frac{1}{16}$.
This perturbed conformal theory describes a single
Ising model
in an external magnetic field $h$ close to, but away from, the critical
temperature
\cite{mccoy,delfino1,zamolodchikov}. 
The coefficient $\tau$ (see (\ref{taueq})) of the
perturbation of conformal dimension
$\frac{1}{2}$ is proportional to the deviation from the
critical temperature, while the coefficient $h$ of the perturbation of
conformal dimension $\frac{1}{16}$ is proportional to the external
magnetic field. The fact that our Georgi-Glashow model, with heavy
fundamental matter, maps onto this dimensionally reduced Ising system
means that we can use the known Ising results to study the nature of the
phase transition in the presence of the heavy fundamental matter field
$\Phi$, whose fugacity
$h$ plays the role of the external magnetic field in the Ising language.

The Ising model with these two perturbations has
been studied extensively \cite{mccoy,delfino1,zamolodchikov}. 
It is believed not
to be exactly soluble, although many exact results are known both at $h=0$
for all
$\tau$
\cite{onsager}, and at $\tau=0$ for all $h$ \cite{magfield,delfino2,caselle}.
Nevertheless, much is known about the system (\ref{lcrit}) with both
perturbations. For example, it has been shown in \cite{zamolodchikov}
that
for the Ising system perturbed by the operators of
conformal dimension
$\frac{1}{2}$ and $\frac{1}{16}$, the free energy can be written as
\begin{eqnarray}F(\tau,h)={2\tau^2\over 15\pi}\log h+
f\left({\tau\over |h|^{8/15}}\right)
\label{result}
\end{eqnarray}
where the function $f(x)$ on the RHS is an {\it analytic}
function for all real
$x$, including $x=0$. This is a highly nontrivial result, which has not
yet been proved rigorously, but which is strongly supported by  numerical
results \cite{zamolodchikov}, as well as by the exact results available
from the $\tau=0$ and the $h=0$ limits. The analyticity relation
(\ref{result}) is a very significant and powerful result.  The
analyticity
of the function $f(x)$ means that the theory has no phase transitions.
This
implies that the second order Ising transition at $M^2\to\infty$ (i.e,
in the theory {\it without} fundamental matter fields) becomes a
crossover
at finite $M^2$. Thus, the second order deconfining phase transition
found
in \cite{gg1} for the finite temperature 2+1 dimensional Georgi-Glashow
model changes into a crossover with the inclusion of fundamental
matter fields that have a heavy but finite mass.

It is also known \cite{zamolodchikov} that away from $\tau=0$,
the free energy has an expansion in powers of
$\alpha={h\over\tau^{15/8}}$.
(This unusual power $15/8$ is a simple consequence of the fact that the
field $h$ has mass dimension $15/8$, while $\tau$ has mass dimension
$1$). Below the transition, where
$\tau<0$, this expansion contains only even powers of
$\alpha$,  while above the transition, where $\tau>0$, it contains both
odd and even powers of $\alpha$. This result is consistent with our
earlier physical estimates, in (\ref{entropybelow}) and
(\ref{entropyabove}), of the behaviour of the entropy on the fugacity
$h$, based on the dipole picture. The width of the crossover region is
determined by the temperature for which the expansion parameter $\alpha$
is small, and therefore
\begin{eqnarray}
T-{g^2\over 4\pi}\propto\zeta^{-1} h^{8\over 15}=
\exp\left(-{32\pi M\over 15 g^2}+{4\pi M_W\over g^2}\right) .
\end{eqnarray}
The increase in entropy which we estimated before in (\ref{entropybelow})
and (\ref{entropyabove}) happens within this range of temperatures. In
particular this tells us that the dependence of the specific heat on $h$
is
\begin{eqnarray}C_V=T{\partial S\over\partial T}\bigg{|}_{V}\propto
\zeta h^{7\over
15}=
\exp\left(-{28\pi M\over 15 g^2}-{4\pi M_W\over g^2}\right) .
\end{eqnarray}To conclude, we note that it would be interesting to
connect our
results with those of \cite{Agasian:2001an} which considers massless
fundamental fermions. We also note that our results can also be
interpreted
in the framework of the $Z_2$ gauge theory. As shown  in \cite{fosco},
the
dual description of the Georgi-Glashow model with heavy fundamental
matter
is a local $Z_2$ gauge theory with matter fields  at weak coupling. The
$Z_2$ gauge  coupling constant  is related to the mass of the fundamental
fields as $e^2\propto M^{-1}$. Thus our results predict that the hot
$Z_2$
theory with matter does not undergo a phase transition but rather a sharp
crossover, with the width of the crossover region nonperturbatively small
at weak coupling
$\Delta T\propto\exp(-{aT_c\over e^2})$.

\begin{acknowledgments}
We thank I. Kogan, M. Ogilvie and B. Tekin for useful discussions,
and A. Tsvelik for pointing out reference \cite{zamolodchikov} to us.
GD and SN are supported  by the U.S. DOE grant DE-FG02-92ER40716.
AK is supported by PPARC, and thanks the University of Connecticut for
the
support of a Guest Professorship Award.
\end{acknowledgments}

\end{document}